\documentclass[twocolumn,pra,showpacs,superscriptaddress]{revtex4-1}
\usepackage{amssymb}
\usepackage{amsmath}
\usepackage{graphicx}
\usepackage{epsfig}

\begin{document}

\title{Asymmetric transmission through a flux-controlled non-Hermitian
scattering center}
\author{X. Q. Li}
\affiliation{School of Physics, Nankai University, Tianjin 300071, China}
\author{X. Z. Zhang}
\affiliation{College of Physics and Materials Science, Tianjin Normal University, Tianjin 300387, China}
\author{G. Zhang}
\affiliation{School of Physics, Nankai University, Tianjin 300071, China}
\author{Z. Song}
\email{songtc@nankai.edu.cn}
\affiliation{School of Physics, Nankai University, Tianjin 300071, China}

\begin{abstract}
We study the possibility of asymmetric transmission induced by a
non-Hermitian scattering center embedded in a one-dimensional waveguide,
motivated by the aim of realizing quantum diode in a non-Hermitian system.
It is shown that a $\mathcal{PT}$ symmetric non-Hermitian scattering center
always has symmetric transmission although the dynamics within the isolated
center can be unidirectional, especially at its exceptional point. We
propose a concrete scheme based on a flux-controlled non-Hermitian
scattering center, which comprises a non-Hermitian triangular ring threaded
by an Aharonov-Bohm flux. The analytical solution shows that such a complex
scattering center acts as a diode at the resonant energy level of the
spectral singularity, exhibiting perfect unidirectionality of the
transmission. The connections between the phenomena of the asymmetric
transmission and reflectionless absorption are also discussed.
\end{abstract}

\maketitle


\section{Introduction}

\label{introduction}

Asymmetric transmission is of significant interest in the quantum analogues
of electronic devices, such as quantum diode device, which is the key to
quantum information processing in integrated circuits\textbf{\ }\cite{Haus}%
\textbf{.} It is characterized by the non-reciprocal particle transport
along the opposite directions. Recently, it has been reported that the
unidirectional transport can be realized in practicle systems\textbf{\ }\cite%
{FanSH1,Onchip,Silicon,FanSH2}.\ A non-Hermitian Hamiltonian can possess
peculiar features that have no Hermitian counterpart. A typical one is the
non-reciprocal dynamics, which has been observed in experiments \cite%
{Observe}. However, it has not been paid due attention by the physics
community until the discovery of non-Hermitian Hamiltonians with parity-time
symmetry, which have a real spectrum \cite{Bender}. It has boosted the
research on the complex extension of quantum mechanics on a fundamental
level \cite{Ann,JMP1,JPA1,JPA2,PRL1,JMP2,JMP3,JMP4,JPA3,JPA4,JPA5}.
Recently, the concept of spectral singularity of a non-Hermitian system has
gained a lot of attention \cite{PRA1,PRB1,Ali3,PRA3,JMP5,PRD1,PRA4,PRA5,PRA6}%
, motivated by the possible physical relevance of this since the pioneer
work of\ Mostafazadeh \cite{PRL3}.\ The majority of previous works focus on
the non-Hermitian system in the absence of an external magnetic field \cite%
{PRA2,JPA6,Ali3,PRA13,prd2,prd3,prd4,prd5,prd6,prd7,prd8,AnnZXZ}.

The aim of this work is to study the possibility of asymmetric transmission
induced by a non-Hermitian scattering center embedded in a one-dimensional
waveguide, motivated by the recent investigation on the physical relevance
of a spectral singularity. It is shown that a $\mathcal{PT}$-symmetric
non-Hermitian scattering center always has symmetric transmission although
the non-reciprocal\textbf{\ }dynamics within the isolated center is allowed,
especially at its exceptional point \cite{JL,ZXZ}. We consider a $\mathcal{PF%
}$-symmetric non-Hermitian scattering center, which comprises a
non-Hermitian triangular ring threaded by an Aharonov-Bohm flux, where $%
\mathcal{F}$\ is the action of flipping the\textbf{\ }flux. We show that
such a complex scattering center acts as a diode, which is characterized by
the different performances of transmission coefficients along the opposite
directions.\textbf{\ }Furthermore, it is found that the perfect
unidirectionality of the transmission is a signature of the existence of a
spectral singularity. And the criterion for spectral singularity by transfer
matrix is not applicable to the present system due to the presence of the
magnetic field.

This paper is organized as follows. In Section \ref{Symmetric transmission},
we present a general formalism for the scattering problem. In Section \ref%
{Perfect unidirectionality}, the Hamiltonian for asymmetric transmission is
constructed and the analytical scattering solution is obtained. In Section %
\ref{Spectral singularity}, we study the connection between the perfect
unidirectionality and the spectral singularity. Finally, we give a summary
and discussion in Section \ref{Summary and discussion}.

\section{Symmetric transmission}

\label{Symmetric transmission}In this section, we present a general
formalism for one-dimensional scattering process of several types of
scattering centers\cite{Muga,AnnPhys}. We will show that the asymmetric
transmission,\textbf{\ }which is the base of a quantum diode, cannot be
realized via $\mathcal{P}$, $\mathcal{T}$, and $\mathcal{PT}$-symmetric%
\textbf{\ }non-Hermitian scattering centers. However, it may be possible for
the non-Hermitian scattering center with a internal degree of freedom.

\begin{figure}[tbp]
\includegraphics[ bb=44 140 556 789, width=7.0 cm, clip]{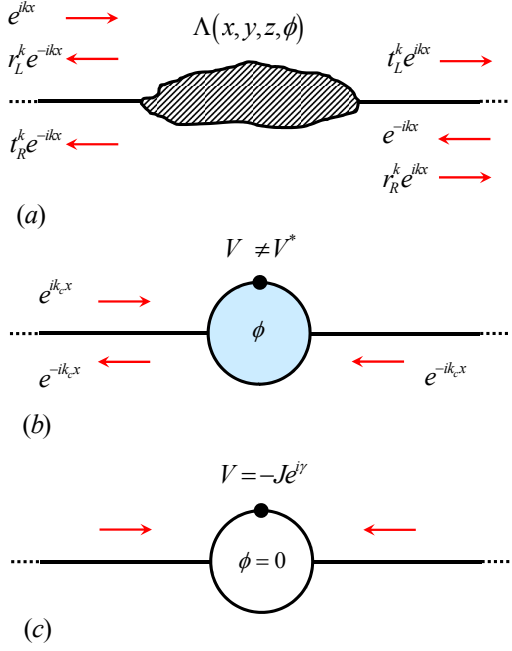}
\caption{(Color online) Sketch of one-dimensional scattering systems for
incident waves from the left and right. (a) An arbitrary scattering center
with $3$-dimensional structure and magnetic flux, is represented by $\Lambda
(x,y,z,\protect\phi )$, which can be non-Hermitian or possesses certain
symmetry. (b) A tight-binding scattering center which comprises a
non-Hermitian triangular ring threaded by an Aharonov-Bohm flux $\protect%
\phi $. The non-Hermiticity arises from on-site complex potential $V$. It
turns out that the optimal $\protect\phi $\ and $V$\ can leads to perfect
asymmetric transmission, the diode characteristic: $r_{\mathrm{R,L}}^{k}=t_{%
\mathrm{L}}^{k}=0$, $\left\vert t_{\mathrm{R}}^{k_{{}}}\right\vert =1$.}
\label{fig1}
\end{figure}

This term \textquotedblleft one-dimensional\textquotedblright\ refers to the
space domain of incident, reflected and transmitted waves, rather than that
of scattering center. This constraint requires the asymptotic eigenfunctions
to be one-dimensional plane waves. Consider a scattering problem for an
arbitrary scattering center, which is schematically illustrated in Fig. \ref%
{fig1}(a). According to the above analysis, for the left and right incident
waves, we have%
\begin{equation}
\psi _{\mathrm{L}}^{k}\left( x\right) =\left\{
\begin{array}{cc}
e^{ikx}+r_{\mathrm{L}}^{k}e^{-ikx}, & \left( x\ll 0\right) \\
t_{\mathrm{L}}^{k}e^{ikx}, & \left( x\gg 0\right)%
\end{array}%
\right. ,  \label{psi_L}
\end{equation}%
and%
\begin{equation}
\psi _{\mathrm{R}}^{k}\left( x\right) =\left\{
\begin{array}{cc}
t_{\mathrm{R}}^{k}e^{-ikx}, & \left( x\ll 0\right) \\
e^{-ikx}+r_{\mathrm{R}}^{k}e^{ikx}, & \left( x\gg 0\right)%
\end{array}%
\right. .  \label{psi_R}
\end{equation}%
Combining $\psi _{\mathrm{R}}^{k}\left( x\right) $\ and $\psi _{\mathrm{L}%
}^{-k}\left( x\right) $ into the form $t_{\mathrm{L}}^{-k}\psi _{\mathrm{R}%
}^{k}\left( x\right) -\psi _{\mathrm{L}}^{-k}\left( x\right) $\ and
comparing it to $-r_{\mathrm{L}}^{-k}\psi _{\mathrm{L}}^{k}\left( x\right) $%
, we have the following relations for the reflection and transmission
amplitudes

\begin{eqnarray}
t_{\mathrm{L}}^{-k}t_{\mathrm{R}}^{k}+r_{\mathrm{L}}^{-k}r_{\mathrm{L}}^{k}
&=&1,  \label{rt1} \\
t_{\mathrm{L}}^{-k}r_{\mathrm{R}}^{k}+r_{\mathrm{L}}^{-k}t_{\mathrm{L}}^{k}
&=&0.  \label{rt2}
\end{eqnarray}%
It still holds if we take $\mathrm{L}\leftrightarrows \mathrm{R}$\ or\ $%
k\longrightarrow -k$.

In the following, we investigate the symmetry of the transmission for
several types of scattering centers\textbf{.} For a scattering center with%
\textbf{\ }time reversal\textbf{\ }$\mathbf{(}\mathcal{T})$\textbf{\ }%
symmetry, where time-reversal operator $\mathcal{T}$\ has the function $%
\mathcal{T}i\mathcal{T}^{-1}=-i$,\ the $\mathcal{T}$ symmetry brings up
additional constraints for the coefficients. One can obtain the following
relations%
\begin{eqnarray}
\left( t_{\mathrm{L}}^{k}\right) ^{\ast }t_{\mathrm{R}}^{k}+\left( r_{%
\mathrm{L}}^{k}\right) ^{\ast }r_{\mathrm{L}}^{k} &=&1,  \label{C_rt1} \\
\left( t_{\mathrm{L}}^{k}\right) ^{\ast }r_{\mathrm{R}}^{k}+\left( r_{%
\mathrm{L}}^{k}\right) ^{\ast }t_{\mathrm{L}}^{k} &=&0,  \label{C_rt2}
\end{eqnarray}%
due to the fact that the conjugations of Eq. (\ref{psi_L}) and (\ref{psi_R})
are still the asymptotic eigenfunctions of the system. Together with the
continuity of probability currents
\begin{equation}
\left\vert t_{\mathrm{L,R}}^{k}\right\vert ^{2}+\left\vert r_{\mathrm{L,R}%
}^{k}\right\vert ^{2}=1,  \label{continuity}
\end{equation}%
we have the symmetry relations%
\begin{equation}
t_{\mathrm{R}}^{k}=t_{\mathrm{L}}^{k}\text{, }r_{\mathrm{R}}^{k}=r_{\mathrm{L%
}}^{k}.  \label{symmetry}
\end{equation}%
It indicates that for a Hermitian scattering center, the asymmetry potential
cannot lead to transmission asymmetry.

A natural question is whether a non-Hermitian scattering center can lead to
the asymmetrical transmission. Before we answer this question we would like
to point that serval types of non-Hermitian scattering center cannot be the
candidate. This may provide a guidance for the diode design. To begin with,
a parity symmetric non-Hermitian scattering center should exhibit symmetric
reflection and transmission. Here the parity operator $\mathcal{P}$\ has the
function $\mathcal{P}x\mathcal{P}^{-1}=-x$. In addition to that, we will
show that a $\mathcal{PT}$-symmetric non-Hermitian scattering center also
possesses the transmission symmetry. For a $\mathcal{PT}$-symmetric
scattering center, wave functions obtained by the $\mathcal{PT}$ action on
the Eqs. (\ref{psi_L}) and (\ref{psi_R}) are still the asymptotic
eigenfunctions.

Comparing $\mathcal{PT}\psi _{\mathrm{R}}^{k}\left( x\right) \left( \mathcal{%
PT}\right) ^{-1}$ and $\psi _{\mathrm{L}}^{-k}\left( x\right) $, we have%
\begin{equation}
\left( r_{\mathrm{R}}^{k}\right) ^{\ast }=r_{\mathrm{L}}^{-k}\text{, }\left(
t_{\mathrm{R}}^{k}\right) ^{\ast }=t_{\mathrm{L}}^{-k},
\end{equation}%
which still holds if we take $\mathrm{L}\leftrightarrows \mathrm{R}$\ or\ $%
k\longrightarrow -k$. Together with Eq. (\ref{rt1}), we have\
\begin{equation}
\left( t_{\mathrm{R}}^{k}\right) ^{\ast }t_{\mathrm{R}}^{k}+\left( r_{%
\mathrm{R}}^{k}\right) ^{\ast }r_{\mathrm{L}}^{k}=1,
\end{equation}%
and%
\begin{equation}
\left( t_{\mathrm{L}}^{k}\right) ^{\ast }t_{\mathrm{L}}^{k}+\left( r_{%
\mathrm{R}}^{k}\right) ^{\ast }r_{\mathrm{L}}^{k}=1,
\end{equation}%
which lead to%
\begin{equation}
\left\vert t_{\mathrm{R}}^{k}\right\vert =\left\vert t_{\mathrm{L}%
}^{k}\right\vert .
\end{equation}%
This result indicates that it is impossible to construct a diode, a
scattering center allowing unidirectional flow, by a $\mathcal{PT}$%
-symmetric non-Hermitian scattering center in the framework of this paper,
although it cannot tell us which type of structure meets the demand.

Now we consider the case where the scattering center has an internal degree
of freedom $\phi $\ as illustrated in Fig. \ref{fig1} (a). The Hamiltonian
has the $\mathcal{PF}$-symmetry, i.e.,
\begin{equation}
\mathcal{PF}H\left( \mathcal{PF}\right) ^{-1}=H,  \label{PF}
\end{equation}%
where $\mathcal{F}$\ is the $\phi $-flip operator, defined as $\mathcal{F}%
H\left( \phi \right) \mathcal{F}^{-1}=H\left( -\phi \right) $. Applying the $%
\mathcal{PF}$-operator on the Eqs. (\ref{psi_L}) and (\ref{psi_R}), we
obtain the new solutions of the Hamiltonian, which lead to the relations

\begin{equation}
t_{\mathrm{L}}^{k}\left( \phi \right) =t_{\mathrm{R}}^{k}\left( -\phi
\right) \text{, }r_{\mathrm{L}}^{k}\left( \phi \right) =r_{\mathrm{R}%
}^{k}\left( -\phi \right) .  \label{phi-phi}
\end{equation}%
In contrast to the systems with $\mathcal{P}$, $\mathcal{T}$, and $\mathcal{%
PT}$-symmetry respectively, one cannot get the conclusion of the\ symmetric
transmission. It opens the possibility of $t_{L}^{k}\left( \phi \right) \neq
t_{R}^{k}\left( \phi \right) $, and a perfect diode for the specific $k_{c}$
and $\phi _{c}$, i.e.,%
\begin{equation}
\left\vert t_{\mathrm{R}}^{k_{c}}\left( \phi _{c}\right) \right\vert ^{2}=1%
\text{, }t_{\mathrm{L}}^{k_{c}}\left( \phi _{c}\right) =0\text{, }r_{\mathrm{%
R,L}}^{k_{c}}\left( \phi _{c}\right) =0,  \label{perfect diode}
\end{equation}%
does not contradict the general constraint relations in Eq. (\ref{phi-phi}).
\begin{figure}[tbp]
\includegraphics[ bb=64 116 492 720, width=7.0 cm, clip]{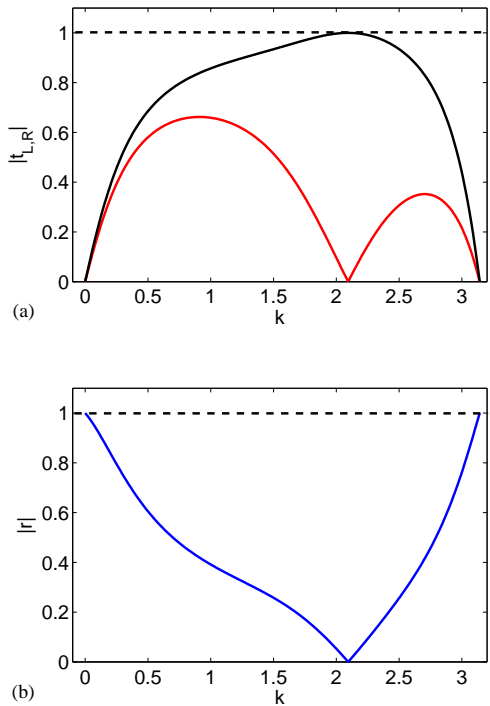}
\caption{(Color online) The absolute value of the transmission (a) and
reflection (b) amplitude profiles as functions of the wave number $k$. It
shows the perfect asymmetric behavior at $k_{c}=2\protect\pi /3$.}
\label{fig2}
\end{figure}
In the following section, we propose a concrete example, a non-Hermitian $%
\mathcal{PF}$-symmetric scattering center embedded in a one-dimensional
tight-binding network, which exhibits perfect unidirectionality.

\section{Perfect unidirectionality}

\label{Perfect unidirectionality}Inspired by the previous work \cite{YS} and
the analysis in above section, we start our design of diode scattering
center by a simplest geometry, a three-site cluster. The threading flux and
single on-site complex potential can destroy the $\mathcal{P}$ symmetry as
well as the $\mathcal{PT}$ symmetry. However, it possesses $\mathcal{PF}$%
-symmetry with the internal degree of freedom being the threading flux.

The Hamiltonian of the concerned scattering tight-binding network has the
form%
\begin{equation}
H=H_{\mathrm{L}}+H_{\mathrm{R}}+H_{\mathrm{Tri}},  \label{H}
\end{equation}%
where%
\begin{eqnarray}
H_{\mathrm{L}} &=&-J\sum_{j=-\infty }^{-2}(a_{j+1}^{\dagger }a_{j}+\mathrm{%
H.c}),  \label{H_L} \\
H_{\mathrm{R}} &=&-J\sum_{j=1}^{\infty }(a_{j+1}^{\dagger }a_{j}+\mathrm{H.c}%
),  \label{H_R}
\end{eqnarray}%
represent the left (HL) and right (HR) waveguides with real $J$ and%
\begin{equation}
H_{\mathrm{Tri}}=-e^{i\phi /3}J\left( a_{0}^{\dagger }a_{-1}+a_{1}^{\dagger
}a_{0}+a_{-1}^{\dagger }a_{1}\right) +\mathrm{H.c}+Va_{0}^{\dagger }a_{0},
\label{H_T}
\end{equation}%
describes a non-Hermitian scattering center, with the non-Hermiticity
arising from the complex potential $V\neq V^{\ast }$. It is a triangular
lattice threaded by a magnetic flux $\phi $, which satisfies the Eq. (\ref%
{PF}) and is schematically illustrated in Fig. \ref{fig1}(b). We note that
zero $\phi $ leads to $\mathcal{P}$ symmetry, real $V$ leads to $\mathcal{PT}
$ symmetry, both of which are the obstacles for the transmission asymmetry
as shown in above section. In this paper, we consider the case with complex
potential $V=-Je^{i\gamma }$ ($\gamma \in \left( 0,\pi \right) $).

We consider the left and right incident scattering processes. We focus our
study on single-particle subspace spanned by the basis $\left\{ \left\vert
j\right\rangle =a_{j}^{\dagger }\left\vert 0\right\rangle \right\} $.\ The
discrete version of Eqs. (\ref{psi_L}) and (\ref{psi_R}) has the form%
\begin{equation}
\psi _{\mathrm{L}}^{k}\left( j\right) =\left\{
\begin{array}{cc}
e^{ikj}+r_{\mathrm{L}}^{k}e^{-ikj}, & \left( j\ll 0\right) \\
t_{\mathrm{L}}^{k}e^{ikj}, & \left( j\gg 0\right)%
\end{array}%
\right. ,  \label{Psi_Lj}
\end{equation}%
and%
\begin{equation}
\psi _{\mathrm{R}}^{k}\left( j\right) =\left\{
\begin{array}{cc}
t_{\mathrm{R}}^{k}e^{-ikj}, & \left( j\ll 0\right) \\
e^{-ikj}+r_{\mathrm{R}}^{k}e^{ikj}, & \left( j\gg 0\right)%
\end{array}%
\right. ,  \label{Psi_Rj}
\end{equation}%
which correspond to the Bethe Ansatz wave function. Employing Bethe Ansatz
technique, we have%
\begin{equation}
r_{\mathrm{L}}^{k}=r_{\mathrm{R}}^{k}=-\frac{\cos \phi +\cos k}{\Omega
\left( k,\phi ,\gamma \right) },  \label{R_LR}
\end{equation}%
\begin{equation}
t_{\mathrm{L}}^{k}=\frac{ie^{-i\phi /3}\sin k\left( e^{i\phi }+2\cos
k-e^{i\gamma }\right) }{\Omega \left( k,\phi ,\gamma \right) },
\end{equation}%
\begin{equation}
t_{\mathrm{R}}^{k}=t_{\mathrm{L}}^{k}\left( \phi \rightarrow -\phi \right) ,
\end{equation}%
with
\begin{equation}
\Omega \left( k,\phi ,\gamma \right) =e^{ik}\left[ i\sin k\left( 2\cos
k-e^{i\gamma }\right) +e^{ik}\cos \phi +1\right] .
\end{equation}%
For an incident wave with momentum $k_{c}=\gamma $\ from left or right, from
Eq. (\ref{R_LR}) we have
\begin{eqnarray}
r_{\mathrm{L}}^{k_{c}} &=&r_{\mathrm{R}}^{k_{c}}=0,  \label{diode RT} \\
t_{\mathrm{R}}^{k_{c}} &=&e^{i\pi /3}e^{-i4\gamma /3}\text{, }t_{\mathrm{L}%
}^{k_{c}}=0,  \notag
\end{eqnarray}%
which exhibits perfect unidirectionality, when the flux $\phi $\ takes the
value $\phi _{c}=\pi -\gamma $. In order to illustrate the asymmetric
transmission effect we consider the case with $\gamma =2\pi /3$ and $\phi
=\pi /3$. Fig. \ref{fig2} shows transmission and reflection profiles as a
function of the wave number $k$. The perfect asymmetric behavior with $%
\left\vert r_{\mathrm{L,R}}^{k_{c}}\right\vert =\left\vert t_{\mathrm{L}%
}^{k_{c}}\right\vert =0.0$ and $\left\vert t_{\mathrm{R}}^{k_{c}}\right\vert
=1.0$ at $k_{c}=2\pi /3$, as expected, is observed. It shows that there is a
relative wider region around $k_{c}$, within which the system still exhibits
the diode characteristic approximately.

\begin{figure}[tbp]
\includegraphics[ bb=66 122 508 724, width=7.0 cm, clip]{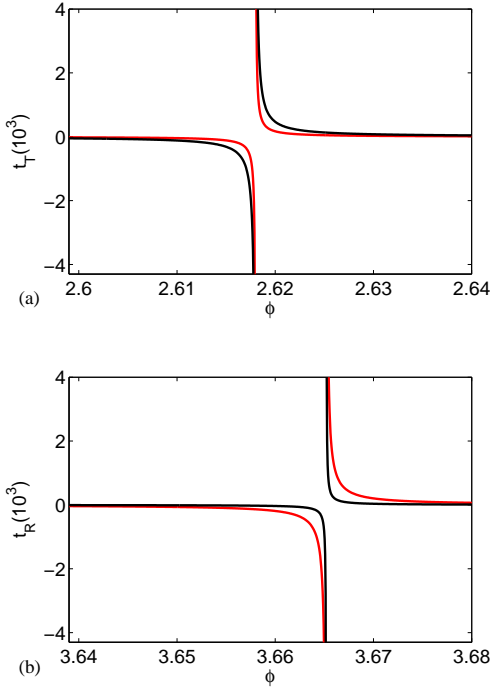}
\caption{(Color online) Plots of the transmission amplitudes from Eq. (%
\protect\ref{t_dagger}) with $k=\protect\gamma =\protect\pi /6$. It shows
that real and imaginary parts of the amplitudes are discontinuous and
divergent at $\protect\phi =5\protect\pi /6$ and $\protect\phi =7\protect\pi %
/6$, respectively. }
\label{fig3}
\end{figure}
\section{Spectral singularity}

\label{Spectral singularity}In this section, we will show that the
occurrence of perfect unidirectionality\ is related to the presence of a
spectral singularity. To this end, we consider the solution of the
Hamiltonian%
\begin{equation}
H^{\dag }=H_{\mathrm{L}}+H_{\mathrm{R}}+H_{\mathrm{Tri}}^{\dag }  \label{C_H}
\end{equation}%
where%
\begin{eqnarray}
H_{\mathrm{Tri}}^{\dag } &=&-e^{i\phi /3}J\left( a_{0}^{\dagger
}a_{-1}+a_{1}^{\dagger }a_{0}+a_{-1}^{\dagger }a_{1}\right)  \label{C_H_T} \\
&&+\mathrm{H.c}-Je^{-i\gamma }a_{0}^{\dagger }a_{0},  \notag
\end{eqnarray}%
which is Hermitian conjugation of $H$. According to the pseudo-Hermitian
quantum mechanics \cite{Ali2}, The eigenfunctions of $H$\ and $H^{\dag }$\
can construct the complete biorthogonal basis except the case of spectral
singularity, at which the complete biorthonormal set is spoiled. By the same
procedure, the scattering wavefunctions can be obtained in the form

\begin{equation}
\overline{\psi }_{\mathrm{L}}^{k}\left( j\right) =\left\{
\begin{array}{cc}
e^{ikj}+\overline{r}_{\mathrm{L}}^{k}e^{-ikj}, & \left( j\ll 0\right) \\
\overline{t}_{\mathrm{L}}^{k}e^{ikl}, & \left( j\gg 0\right)%
\end{array}%
\right. ,  \label{psiL_dagger}
\end{equation}%
and%
\begin{equation}
\overline{\psi }_{\mathrm{R}}^{k}\left( j\right) =\left\{
\begin{array}{cc}
\overline{t}_{\mathrm{R}}^{k}e^{-ikj}, & \left( j\ll 0\right) \\
e^{-ikj}+\overline{r}_{\mathrm{R}}^{k}e^{ikj}, & \left( j\gg 0\right)%
\end{array}%
\right. ,  \label{psiR_dagger}
\end{equation}%
where%
\begin{eqnarray}
\overline{r}_{\mathrm{L}}^{k} &=&\overline{r}_{\mathrm{R}}^{k}=r_{\mathrm{L}%
}^{k}\left( \gamma \rightarrow -\gamma \right) ,  \label{r_dagger} \\
\overline{t}_{\mathrm{L,R}}^{k} &=&t_{\mathrm{L,R}}^{k}\left( \gamma
\rightarrow -\gamma \right) .  \label{t_dagger}
\end{eqnarray}%
However, it becomes a little complicated when we consider the solution for $%
k=k_{c}$. We find that the eigenfunctions with $k_{c}$\ does not exist when
the flux $\phi $\ takes the value $\phi _{c}=\pi -\gamma $. We investigate
the limit of $\overline{r}_{\mathrm{L,R}}^{k}$ and $\overline{t}_{\mathrm{L,R%
}}^{k}$ as $\left( k,\phi \right) \rightarrow \left( \gamma ,\pi -\gamma
\right) $ along the following two paths: (I) $\phi =\pi -k$, $k\rightarrow
\gamma ^{+}$\ and (II) $\phi =\pi -k$, $k\rightarrow \gamma ^{-}$,
respectively. A straightforward calculation shows that two different paths
gives unequal limits, i.e.,%
\begin{equation}
\text{(I, II) }\left\{
\begin{array}{c}
\lim_{k\rightarrow \gamma ^{\pm }}\overline{r}_{\mathrm{L,R}}^{k}=0 \\
\lim_{k\rightarrow \gamma ^{\pm }}\overline{t}_{\mathrm{L}}^{k}=\mp \infty
\pm i\infty \\
\lim_{k\rightarrow \gamma ^{\pm }}\overline{t}_{\mathrm{R}}^{k}=e^{i\pi
/3}e^{-i4\gamma /3}%
\end{array}%
\right. .  \label{(I,II)}
\end{equation}%
We see that the transmission amplitude $\overline{t}_{\mathrm{L}}^{k}$ has a
singularity at the point $k=k_{c}=\gamma $, which indicates that the Bathe
Ansatz solution in Eqs (\ref{psiL_dagger}) and (\ref{psiR_dagger}) do not
exist. Then the complete biorthogonality\ of the eigenfunctions of $H$\ and $%
H^{\dag }$\ is destroyed at this point. We can also investigate this point
from other way, taking the limits of $\overline{r}_{\mathrm{L,R}}^{k}$ and $%
\overline{t}_{\mathrm{L,R}}^{k}$ as $\left( k,\phi \right) \rightarrow
\left( \gamma ,\pi -\gamma \right) $ along the following two paths: (I) $%
k=\gamma $, $\phi \rightarrow \pi -\gamma ^{+}$\ and (II) $k=\gamma $, $\phi
\rightarrow \pi -\gamma ^{+}$, respectively. To demonstrate the singularity,
we plot $t_{\mathrm{L,R}}^{k_{c}}$ as functions of $\phi $ for $k_{c}=\gamma
=\pi /6$ in Fig. \ref{fig3}. The profiles of the plots show clearly that the
one-sided limits from the left and from the right for the real and imaginary
parts are discontinuous and divergent at $\phi =5\pi /6$ and $\phi =7\pi /6$%
, respectively. Therefore we conclude that the perfect asymmetric
transmission corresponds to the existence of the spectral singularity of the
non-Hermitian diode model.

On the other hand, it is noted that the theory of spectral singularity for
non-Hermitian scattering center arising from pure complex potential has been
well established \cite{PRL3,Ali3}. It is shown that the transfer matrix can
be employed to identify the spectral singularity.
\begin{figure}[tbp]
\includegraphics[ bb=40 407 551 785, width=7.0 cm, clip]{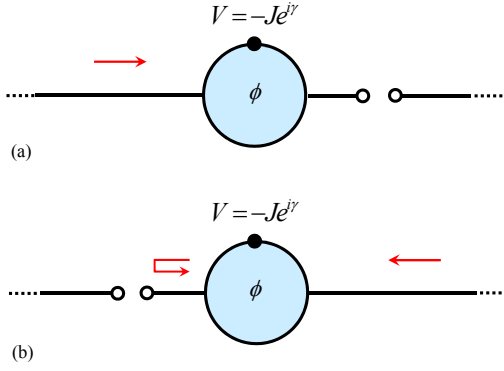}
\caption{(Color online) Sketch of semi-infinite systems possessing
reflectionless absorption characteristic, which can be constructed by
disconnecting the right (a) or left (b) lead from a diode configuration.}
\label{fig4}
\end{figure}
We believe that this formalism is applicable on the discrete system.
Similarly, the eigenvalue equation $H\psi =E\psi $ yields the following
asymptotic expressions for the eigenfunctions of $H$:%
\begin{equation}
\psi _{k}\rightarrow A_{\pm }e^{ikj}+B_{\pm }e^{-ikj}\text{ \ for }%
x\rightarrow \pm \infty .
\end{equation}%
$A_{\pm }$ and $B_{\pm }$ are possibly $k$-dependent complex coefficients
that are related by the so-called transfer matrix $M$ according to%
\begin{equation}
\binom{A_{+}}{B_{+}}=M\binom{A_{-}}{B_{-}}.
\end{equation}%
The transfer matrix reveals almost complete features of the scattering
center. Next, consider the Jost solution of $H$ can be readily constructed
from Eq. (\ref{diode RT}) as the form%
\begin{eqnarray}
\psi _{+}^{k}\left( j\right)  &=&\left\{
\begin{array}{cc}
e^{ikj}/t_{\mathrm{L}}^{k}+r_{\mathrm{L}}^{k}e^{-ikj}/t_{\mathrm{L}}^{k}, &
\left( j\ll 0\right)  \\
e^{ikj}, & \left( j\gg 0\right)
\end{array}%
\right. \text{, } \\
\psi _{-}^{k}\left( j\right)  &=&\left\{
\begin{array}{cc}
e^{-ikj}, & \left( j\ll 0\right)  \\
e^{-ikj}/t_{\mathrm{R}}^{k}+r_{\mathrm{R}}^{k}e^{ikj}/t_{\mathrm{R}}^{k}, &
\left( j\gg 0\right)
\end{array}%
\right. .
\end{eqnarray}%
which satisfy the asymptotic boundary conditions:
\begin{equation}
\psi _{\pm }^{k}\left( j\right) \rightarrow e^{\pm ikj}\text{ as }%
j\rightarrow \pm \infty \text{.}
\end{equation}%
Then the corresponding transfer matrix can be written as%
\begin{equation}
M_{k}=\left(
\begin{array}{cc}
\frac{t_{\mathrm{L}}^{k}t_{\mathrm{R}}^{k}-\left( r_{\mathrm{R}}^{k}\right)
^{2}}{t_{\mathrm{R}}^{k}} & \frac{r_{\mathrm{R}}^{k}}{t_{\mathrm{R}}^{k}} \\
-\frac{r_{\mathrm{R}}^{k}}{t_{\mathrm{R}}^{k}} & \frac{1}{t_{\mathrm{R}}^{k}}%
\end{array}%
\right) ,
\end{equation}%
which connects the asymptotic scattering wavefunctions at $\pm \infty $.

Now we consider the case with $\phi =0$, which represents the non-Hermitian
scattering center arising from complex on-site potentials. Straightforward
derivation shows that when the momentum $k=k_{c}$, with%
\begin{equation}
\sin k_{c}=-\frac{\cos \gamma +1}{2\sin \gamma },  \label{sink_c}
\end{equation}%
we have
\begin{equation}
\left( M_{k_{c}}\right) _{22}=1/t_{\mathrm{R}}^{k_{c}}=0.
\end{equation}%
It identifies a spectral singularity at $k_{c}$, which also corresponds to $%
\left\vert t_{\mathrm{R,L}}^{k_{c}}\right\vert =\left\vert r_{\mathrm{R,L}%
}^{k_{c}}\right\vert =\infty $. This result accords with the theorems
proposed in Ref. \cite{PRL3}. Furthermore, the corresponding eigenfunctions
can be written as

\begin{equation}
\psi _{\pm }^{k_{c}}\left( j\right) =\left\{
\begin{array}{cc}
e^{-ik_{c}j}, & \left( j\ll 0\right) \\
e^{ik_{c}j}, & \left( j\gg 0\right)%
\end{array}%
\right. \text{, }  \label{RA solution}
\end{equation}%
due to the fact

\begin{equation}
\lim_{k\rightarrow k_{c}}r_{\mathrm{L}}^{k}/t_{\mathrm{L}}^{k}=\lim_{k%
\rightarrow k_{c}}r_{\mathrm{R}}^{k}/t_{\mathrm{R}}^{k}=1.
\end{equation}%
Obviously, the physics of the solution in Eq. (\ref{RA solution})
corresponds to the undirectional plane wave, which has been proposed in
Refs. \cite{PRB1,PRA14}. It can be seen from the following analysis. The
group velocities of the incident plane waves from left and right are denoted
as\textbf{\ }$\upsilon _{\mathrm{L}}=\upsilon _{k_{c}}^{-}$\textbf{\ }and%
\textbf{\ }$\upsilon _{\mathrm{R}}=\upsilon _{k_{c}}^{+}$\textbf{, }where%
\begin{equation}
\upsilon _{k_{c}}^{\pm }=\left( \frac{\partial E_{k}}{\partial k}\right)
_{\pm k_{c}}=\pm 2J\sin k_{c}.
\end{equation}%
For the center loss potential $V=-Je^{i\gamma }$\ with $\gamma \in \left(
0,\pi \right) ,$ we have $\sin k_{c}<0$\ according to Eq. (\ref{sink_c}) and
then yields $\upsilon _{\mathrm{L}}>0$, $\upsilon _{\mathrm{R}}<0$. It
indicates that the solution in Eq. (\ref{RA solution}) represents the
current flow from both sides to the center. It corresponds to reflectionless
absorbtion with $\mathcal{P}$ symmetry, which is schematically illustrated
in Fig. \ref{fig1}(c).

On the other hand, when we consider the nonzero $\phi $\ case, we will find
that such a criterion is invalid for the spectral singularity under the
condition of the perfect asymmetric transmission. From Eq. (\ref{diode RT}),
the corresponding transfer matrix can be written as%
\begin{equation}
M_{k_{c}}=\left(
\begin{array}{cc}
0 & 0 \\
0 & e^{i\left( 4k_{c}-\pi \right) /3}%
\end{array}%
\right) ,
\end{equation}%
which indicates $M_{22}\neq 0$. It implies that the criterion for the
existence of spectral singularity is not necessary when the magnetic field
is involved.

Finally, we would like to discuss the asymmetric transmission from other
perspective. It has been turned out that, there is another peculiar
phenomenon, reflectionless absorption, in the semi-infinite non-Hermitian
system \cite{PRB1,PRA14}. We have shown in above section that, such a
phenomenon in its symmetrized version, can occur in the present system with
zero $\phi $. Now we will show the connection between the perfect
unidirectionality and reflectionless absorption. In contrast to the previous
study, we consider the system in the presence of a magnetic flux. One can
see this connection simply by disconnecting one of two leads in the system
with the Hamiltonian in Eq. (\ref{H}). Figure \ref{fig4}(a) and (b)
schematically illustrates this geometry, which are obtained by cutting off
the right and left leads, respectively. Then we can conclude that a perfect
diode scattering center can always be reduced to a setup of reflectionless
absorption. However, the latter is not sufficient to construct a diode
device.

\section{Summary and discussion}

\label{Summary and discussion}In summary, we have studied the possibility of
asymmetric transmission induced by a non-Hermitian scattering center
embedded in a one-dimensional waveguide. We have shown that the
non-Hermiticity of a scattering center is not sufficient for the asymmetric
transmission, while it is forbidden for a Hermitian scattering center. We
have constructed a concrete setup possessing the perfect unidirectionality
of the transmission, which comprises a non-Hermitian triangular ring
threaded by an Aharonov-Bohm flux. It seems to imply that a magnetic flux is
crucial for such a phenomenon. Furthermore, the analytical solution shows
the connection between the perfect unidirectionality and spectral
singularity. We have also showed that the criterion for spectral singularity
associated with the transfer matrix is invalid when the magnetic field is
involved.

\acknowledgments We acknowledge the support of the National Basic Research
Program (973 Program) of China under Grant No. 2012CB921900 and CNSF (Grant
No. 11374163).

\end{document}